\documentclass[aps,prf,superscriptaddress,showpacs,amsmath,amssymb,onecolumn,nofootinbib,longbibliography]{revtex4-2}
\usepackage{amsmath}
\usepackage{amsfonts}
\usepackage{amsthm}
\usepackage{graphics}
\usepackage[draft]{graphicx}
\usepackage{amssymb}
\usepackage{mathtools}
\usepackage{caption,subcaption}
\usepackage{color}
\usepackage{url}
\usepackage[abs]{overpic}
\usepackage{float}
\usepackage[usenames,dvipsnames]{xcolor}
\usepackage[normalem]{ulem}
\usepackage{ragged2e}
\usepackage{tabularx}
\usepackage{booktabs}
\usepackage{soul}
\usepackage{tikz}
\usepackage{diagbox}  % For the diagonal line in the table

\usepackage{natbib}
\usepackage{graphicx}% Include figure files
\usepackage{dcolumn}% Align table columns on decimal point
\usepackage{bm}% bold math
\usepackage[utf8]{inputenc}
\usepackage[T1]{fontenc}
\usepackage[english]{babel}
\usepackage{pgf}
\usepackage[export]{adjustbox}
\usepackage{wrapfig}
\usepackage{xmpmulti}
\usepackage{amsmath}
\usepackage[
  justification=RaggedRight,
  calcwidth=1\linewidth,
]{caption}
\usepackage{hyperref}
\hypersetup{
    colorlinks=true,
    linkcolor=blue,
	citecolor = blue,      
    urlcolor=blue}

\def\kmax{k_{\rm max}}

\newcommand{\disc}{\tikz\fill (0,0) circle (0.85mm);}
\newcommand{\xmarker}{\tikz{\draw[thick] (-0.85mm,-0.85mm) -- (0.85mm,0.85mm); \draw[thick] (-0.85mm,0.85mm) -- (0.85mm,-0.85mm);}}

\def\mbfx{\mathbf{x}}
\def\mbfr{\mathbf{r}}
\def\mbfu{\mathbf{u}}

\def\mbff{\mathbf{f}}
\def\mbfp{\mathbf{p}}

\graphicspath{{../}}

\begin{document}
	
	\title{Orientation Dynamics of Gyrotactic Microswimmers in Turbulent Flows}

	\author{Suraj Kumar Nayak}
	\email{surajkumarnayak96@gmail.com}
	\affiliation{Department of Physics, Indian Institute of Technology Kharagpur, West Bengal - 721 302, India}%
	
	\author{Vishwanath Shukla}
	\email{vishwanath.shukla@phy.iitkgp.ac.in}
	\affiliation{Department of Physics, Indian Institute of Technology Kharagpur, West Bengal - 721 302, India}%
	
	\author{Akshay Bhatnagar}
	\email{akshayphy@gmail.com}
	\affiliation{Department of Physics, Indian Institute of Technology Roorkee, Uttarakhand - 247 667, India}%
	
	\date{\today}
	\begin{abstract} 		
We study the dynamics of gyrotactic microswimmers suspended in
homogeneous and isotropic turbulence by using direct numerical
simulations (DNS). The swimmers are characterized by three
non-dimensional parameters: their aspect ratio ($\gamma$), a
dimensionless swimming speed ($\phi$), and a dimensionless
reorientation time ($\psi$). Strong gyrotaxis (smaller $\psi$) promotes vertical alignment of the swimmers, while weak gyrotaxis leads to nearly isotropic orientations. At low swimming numbers, the orientation distribution is largely shape-independent with spheres and spheroids showing marginally greater vertical alignment than rods, whereas at higher activity the peaks of the distributions exhibit largely shape-independent behavior and the tails show a clear dependence on particle shape. However, at large $\psi$ rods exhibit a stronger alignment along the vertical. We observe that at small $\psi$ the rod-shaped swimmers respond to shear by aligning with the stretching direction of the strain-rate tensor, while at large $\psi$ the alignment with the vorticity vector is preferred. The orientation autocorrelation is found to decay exponentially, with a decay rate that scales as $1/(2\psi)$. Analysis of the mean-squared displacement (MSD) reveals a transition from a ballistic motion at short times to a diffusive regime at longer times. To assess the efficiency of vertical migration, we compute the probability distributions of vertical displacement over a fixed time interval and the time taken to migrate a specific vertical distance.  Furthermore, we use a simplified two-dimensional model for spherical swimmers that qualitatively reproduces the key trends observed in the full three-dimensional (3D) simulations.

	\end{abstract}

\maketitle

\section{Introduction}

Marine and aquatic microorganisms often navigate through turbulent fluid flows for food, reproduction and survival. Many of these microswimmers actively adjust their swimming behaviour in response to local hydrodynamic cues, such as shear and vorticity~\cite{lampert1989adaptive, durham2013turbulence, de2014turbulent, rusconi2015microbes, gustavsson2016preferential, sengupta2017phytoplankton, gupta2020flocking, alageshan2020machine, si2022preferential, qiu2022review}. An interesting class of microswimmers with asymmetrical mass distribution within their body exhibit an inherent directional bias in their motion. This directed swimming, known as gyrotaxis, arises from a subtle interplay between the viscous torque exerted by the surrounding fluid and the gravitational torque.

Experimental and numerical studies have shown that the fast-moving swimmers preferentially sample the upwelling regions of the flow, whereas the slow-moving swimmers sample the downwelling regions of the flow \cite{lampert1989adaptive, durham2011gyrotaxis, durham2013turbulence, gustavsson2016preferential, borgnino2018gyrotactic, cencini2019gyrotactic, wheeler2019not, liu2022accumulation, qiu2022review}. For example, phytoplanktons perform diel vertical migration in turbulent environments for sunlight and nutrient uptake \cite{kessler1985hydrodynamic, lampert1989adaptive, kamykowski1998relationships, durham2013turbulence, gustavsson2016preferential, borgnino2018gyrotactic, wheeler2019not}. Also, in high shear regions, the natural alignment of the microswimmers is disrupted and they undergo tumbling motion, leading to the formation of thin layers that can range from a few centimeters to meters in thickness and extend over several kilometers horizontally. These layers can occur in laminar as well as turbulent flows and hinder the sunlight penetration, increase zooplankton and fish mortality, and at times can trigger harmful algal blooms \cite{durham2009disruption, santamaria2014gyrotactic, rusconi2015microbes,  cencini2019gyrotactic}. Also, the gyrotaxis enables these organisms to orient and migrate preferentially in the flow, often leading to striking cluster patterns in turbulent environments \cite{kessler1984gyrotactic, kessler1985hydrodynamic, jeffery1922motion, roberts1970geotaxis, mitchell1990gyrotaxis, durham2011gyrotaxis, guasto2012fluid, rusconi2015microbes, qiu2022review}, which can be fractal \cite{durham2013turbulence, zhan2014accumulation, de2014turbulent, gustavsson2016preferential, borgnino2018gyrotactic}. The clustering depends on the shape, spherical swimmers show stronger clustering compared to the prolate- or rod-like swimmers~\cite{durham2011gyrotaxis, durham2013turbulence, zhan2014accumulation, de2014turbulent,  gustavsson2016preferential, borgnino2018gyrotactic, cencini2019gyrotactic, liu2022accumulation,qiu2022review}. Moreover, their preferential transport and dispersion for foraging and survival in laminar and turbulent flows has been extensively studied~\cite{mitchell1990gyrotaxis, taylor2012trade, croze2013dispersion, sengupta2017phytoplankton, cencini2019gyrotactic, wheeler2019not, si2022preferential, qiu2022review}. However, the orientation statistics of these microswimmers remains relatively less explored. 

Gyrotactic microswimmers have a natural tendency to align vertically as a result of their bottom-heaviness \cite{roberts1970geotaxis, kessler1984gyrotactic, kessler1985hydrodynamic, pedley1987orientation}. This has been examined both experimentally and theoretically for simple flows; for example, gyrotaxis leads to the formation of bioconvection patterns, wherein particles swim upwards to the top and fall as plumes with velocity greater than the flow velocity~\cite{kessler1984gyrotactic, kessler1985hydrodynamic,  pedley1987orientation}. The local equilibrium orientations of these swimmers have been studied in a variety of experimental setups, such as the vertical Poiseuille flow in a pipe, conical sink flows, 2D shear flows, and the wake of a falling sphere \cite{pedley1987orientation}.

The axisymmetric rod-like microscopic tracer particles are known to exhibit strong alignment with the vorticity vector in turbulent flows and a good alignment with the strain-rate-tensor eigenvector associated with the largest eigenvalue~\cite{pumir2011orientation}. In Ref.~\cite{pujara2018rotations, borgnino2019alignment, pujara2021shape}, the alignment of nonspherical active particles in chaotic, moderately turbulent flows was studied, the rod-like active particles were found to preferentially align with the flow velocity. Also, a comparison of gyrotactic and non-gyrotactic swimmers reveals differences in how they align with the vorticity and strain-rate eigenvectors~\cite{zhan2014accumulation}. Moreover, when the swimming velocity is increased, the alignment with the Lagrangian stretching direction is reduced in the presence of a weak gyrotaxis, whereas it is enhanced if the gryotaxis is strong~\cite{liu2022accumulation}. In Ref.~\cite{lewis2003orientation}, the solutions to the Reynolds-averaged equations for the orientation probability distribution function of spheroidal swimmers in the long-time limit were found to be a one-parameter Fisher distribution. 

In this study, we carry out a systematic investigation of the orientation statistics of gyrotactic microswimmers in a turbulent flow using direct numerical simulations (DNS). We consider three different aspect ratios: spheres, spheroids and rods. The orientation distribution is influenced by the flow gradients, swimmer geometry, and the gyrotactic response. We find that the strongly gyrotactic microswimmers predominantly align with the vertical, while weakly gyrotactic swimmers exhibit nearly isotropic orientation distribution, the alignment is more enhanced for the spheres and spheroids compared to the rods. The rod-shaped swimmers are more influenced by the shear in the flow and tend to align with eigenvector corresponding to the largest eigenvalue of the strain-rate tensor. The rate of exponential decay of the orientation autocorrelations depends on the gyrotactic response time. All the swimmers exhibit ballistic and diffusive transport at short and long times, respectively, irrespective of their gyrotactic strength. A two-dimensional reduced model for spherical swimmers is able to capture the key features of the orientation and transport statistics.

The remainder of this paper is structured as follows. In Sec.~II, we discuss the model and the numerical methods, followed by the presentation of our DNS results in Sec.~III. Section~IV contains a discussion of the reduced model. We give our conclusions in Sec.~V.

\section{The Model and Numerical Setup}

We consider a dilute concentration of gyrotactic microswimmers in a turbulent incompressible three-dimensional (3D) fluid flow. The spatio-temporal evolution of the incompressible fluid velocity field $\mbfu(\mbfx,t)$ is governed by the Navier-Stokes equations (NSE)
\begin{subequations}\label{eq:NSE}
	\begin{align} 
		\frac{\partial {\bf u}}{\partial t} + ({\bf u}\cdot {\bf \nabla}){\bf u} &= - {\bf \nabla} p + \nu \nabla^2 {\bf u} + \mathbf{f}\label{eqn:NS},\\
		{\bf \nabla} \cdot {\bf{u}} &=  0,
		\label{eqn:incompressibility}
	\end{align}
\end{subequations}
where $p(\mbfx,t)$ is the pressure field and $\nu$ is the kinematic viscosity. We set the fluid density $\rho=1$ without any loss of generality. The external body force $\mbff(\mbfx,t)$ helps in maintaining a statistically stationary state by injecting energy at large length scales. The injected energy cascades down to the viscous dissipation length scales, characterized by the Kolmogorov scale, $\eta = (\nu^3/\epsilon)^{1/4}$, where $\epsilon$ is the energy dissipation rate. The time scale and the velocity associated with $\eta$ are $\tau_{\eta} = (\nu/\epsilon)^{1/2}$ and $u_{\eta} = (\nu \epsilon)^{1/4}$, respectively. The turbulent flow is characterized by the Taylor-scale Reynolds number $Re_{\lambda} = (u_{rms} \lambda)/ \nu$, where $u_{rms} = \sqrt{\langle |{\bf u}|^2 \rangle/3}$ is the root mean squared velocity and $\lambda = \sqrt{15 \nu u_{rms}^2/\epsilon}$ is the Taylor microscale.

In gyrotactic microswimmers the geometric center and the center of mass do not coincide because of asymmetric mass distribution, as a result these bottom-heavy swimmers have a natural tendency to swim vertically upwards. We consider gyrotactic microswimmers of size smaller than the Kolmogorov length scale and their density to be close to the fluid density. We prescribe following equations for the dynamical evolution of the position $(\mbfx)$ and the orientation $(\mbfp)$ of the microswimmers
\begin{subequations}
	\begin{gather}
	\dot{\mbfx} = \mbfu(\mbfx,t) + v_s \mbfp, \label{eqn:xdot} \\
	\dot{\mbfp} = \frac{\left[\hat{\mathbf{z}} - (\hat{\mathbf{z}}\cdot \mbfp)\mbfp \right]}{2B}  + \frac{\boldsymbol{\omega}
 \times \mbfp}{2}  + \gamma [\mathbf{S}\mbfp - (\mbfp\cdot \mathbf{S}\mbfp)\mbfp], \label{eqn:pdot}
 \end{gather}
\end{subequations}
where $v_s$ is the self-propulsion speed, $B$ is the gyrotactic reorientation time, $|\mbfp|=1$, $\hat{\mathbf{z}}$ is the unit vector along the $z$-axis (the vertical direction) and $\gamma$ is the aspect ratio of the microswimmer. $\boldsymbol{\omega}
=\nabla \times \mbfu$ and $\mathbf{S} = (1/2) \left[\nabla \mbfu + (\nabla \mbfu)^T \right]$ are the fluid vorticity and the symmetric part of the strain rate tensor, respectively, at the swimmer's position.

The first term on the right hand side of Eq.~\eqref{eqn:pdot}, describes the reorientation of the microswimmers towards the vertical direction~\cite{jeffery1922motion}. We define a stability number, $\psi = B/\tau_{\eta}$, as the ratio of the reorientation time towards the z-axis to the Kolmogorov time. The second and the third terms describe the rotation due to the local vorticity and the local strain rate, respectively, experienced by the swimmers due to the flow~\cite{jeffery1922motion}. The latter depends on the shape of the microswimmer, characterized by $\gamma=(l^2 - d^2)/(l^2 + d^2)$, where $l$ is the length of the swimmer in the direction of ${\bf p}$ and $d$ is the width. In our study, we neglect the inertia of the swimmers, the interactions among the swimmers and the back-reaction of swimmers on the flow.

We perform direct numerical simulations (DNS) of the incompressible 3D NSE Eqs.~\eqref{eq:NSE} using a pseudospectral method with $2/3$ de-aliasing rule on a triply periodic domain of length $2\pi$. We use a second-order, exponential Adams-Bashforth scheme for time marching and a constant-energy-forcing scheme to maintain the flow in a steady state. Our DNS runs involve $256^3$ collocation points, the resulting turbulent flow has $Re_{\lambda} \approx 120$ with $\kmax\eta> 1.1$. % thereby ensuring the resolution of the smallest length scales. 
We integrate Eq.~\eqref{eqn:xdot} and Eq.~\eqref{eqn:pdot} for $1$ million swimmers by using a trilinear interpolation to compute the velocity and its gradients at the swimmer's position. We explore the orientation dynamics for a wide range of parameters: the stability number $\psi = (0.5 ,\ 1,\ 10)$ and the swimming number $\phi \equiv v_s/u_{\eta} \in [0:10]$ for three different aspect ratios $\gamma = 0$, $ 0.5$, and  $1$ that correspond to spheres, spheroids and rod-like swimmers, respectively. We initialize the swimmers at uniform random positions when the background turbulent flow has attained a steady state. For the convergence of statistical quantities, we perform the DNS run for several eddy turnover times. Also, to obtain a good estimate of the mean-squared displacement, we use $10^5$ swimmers and integrate the equations over $50$ eddy turnover times. In Tab.~\ref{tab:paramsdns}, we present the summary of the parameters used in our DNS run.
 
\begin{table}[htp!]
\begin{ruledtabular}
\begin{tabular}{lcccccccccccc}  
   \vspace{2pt}
   {N} & {$\delta t$} & {$\nu$} & {$N_p$}& {$k_{max} \eta$} & {$Re_{\lambda}$} & {$\epsilon$} & {$\eta$} & {$\lambda$} & {$\tau_{\eta}$}& {$\tau_{eddy}$} & {$u_{\eta}$} \vspace{0.1cm} \\
   \hline
   \\
    256 & $5 \times10^{-4}$ & $2.5 \times 10^{-3}$ & $10^6$ &1.14 & 120 & 0.5 & 0.01 & 0.29 & 0.07 & 2.19 & 0.19
\end{tabular}
\end{ruledtabular}
\caption{Table of parameters for the DNS run with $N^3$ collocation
points. $\delta t$ is the time step, $\nu$ is the kinematic viscosity,
$N_p$ is the number of microswimmers, $\epsilon$ is the mean rate of
energy dissipation. $\eta = (\nu^3/\epsilon)^{1/4}$, \ $\tau_{\eta} =
(\nu/\epsilon)^{1/2}$ and $u_{\eta} = (\nu \epsilon)^{1/4}$ are the
Kolmogorov dissipation length scale, time scale and velocity
respectively. $\lambda = \sqrt{15 \nu u_{rms}^2/\epsilon}$ is the
Taylor microscale, $Re_{\lambda} = (u_{rms} \lambda)/ \nu$ is the
Reynolds number based on $\lambda$ and $\tau_{eddy} = \left
(\sum_k{\frac{E(k)/k}{E}} \right)/u_{rms}$ is the large eddy turnover
time, where $E(k)$ is the energy spectrum of the flow, $E$ is the
total energy and $u_{rms}$ is the root-mean-squared velocity of the
flow.}
\label{tab:paramsdns}
\end{table}

\section{Results}

\subsection{Orientation dynamics}

\begin{figure}
	\includegraphics[width = \linewidth]{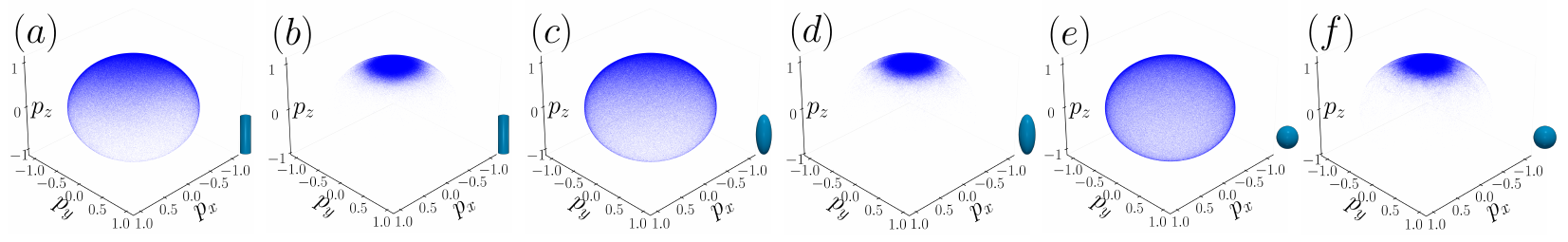}
	\caption{Orientation distribution in $p_x p_y p_z$ space. Rod-like microswimmers for (a) $\psi=10$ and (b) $\psi=0.5$. Spheroids for (c) $\psi=10$ and (d) $\psi=0.5$. Spheres for (e) $\psi=10$ and (f) $\psi=0.5$. Swimming number is kept fixed at $\phi=10$. Projections of the orientation vector along $\hat{\mathbf{x}}$,  $\hat{\mathbf{y}}$ and $\hat{\mathbf{z}}$ are given by $p_x\equiv \mbfp \cdot \hat{\mathbf{x}}$, $p_y\equiv \mbfp \cdot \hat{\mathbf{y}}$ and $p_z\equiv \mbfp \cdot \hat{\mathbf{z}}$, respectively.}
	\label{fig:pxpypz_distri}
\end{figure}

\begin{figure*}
	\includegraphics[width = \linewidth]{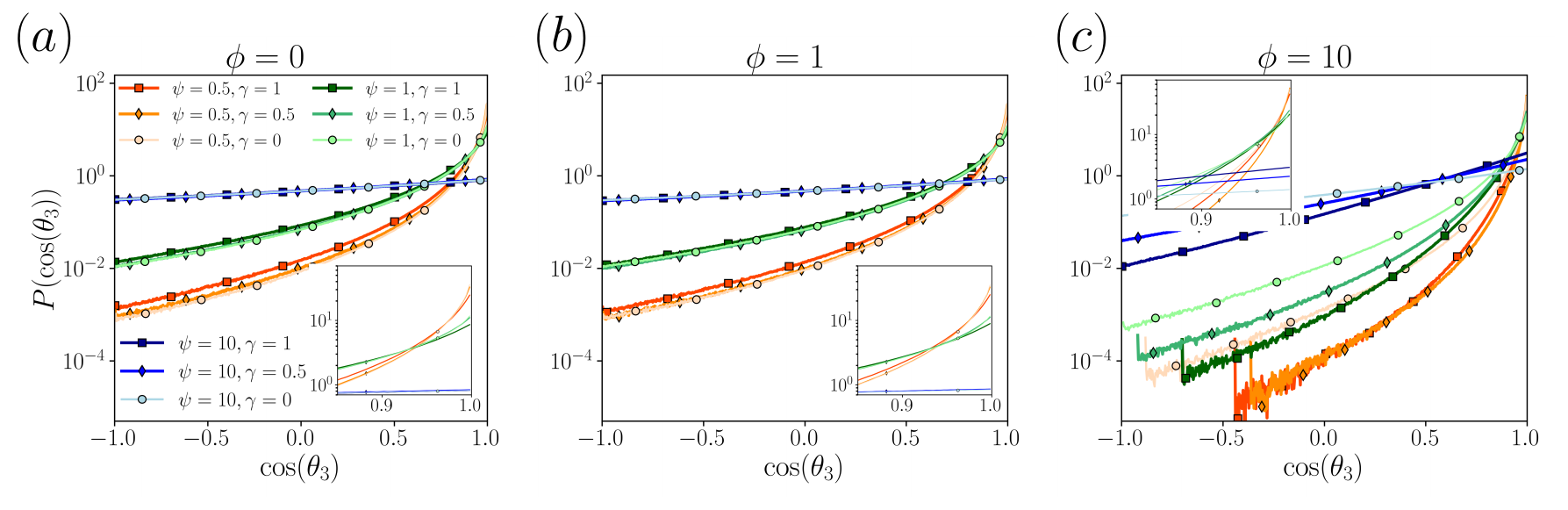}	
	\caption{Probability distribution functions of $\cos \theta_3 (\equiv \mathbf{p}\cdot \hat{\mathbf{z}})$ for swimming numbers: (a) $\phi = 0$, (b) $\phi = 1$, and (c) $\phi = 10$. 
	Behavior is shown for the stability numbers $\psi=0.5$ (orange curves), $1$ (green curves) and $10$ (blue curves). Each $(\phi,\psi)$ combination is explored for three shapes: rods ($\gamma=1$, square markers), spheroids ($\gamma=0.5$, diamond markers) and spheres ($\gamma=0$, circle markers). The inset shows the magnified part of the peaks of the distributions in the main figure. The color and marker representations indicated in the legend of the first plot are common to all plots in the figure.}
	\label{fig:pdf_pz}
\end{figure*}

The projection of the orientation vector $\mbfp$ along the $x$-, $y$-
and $z$- directions are given by $p_x\equiv \mbfp \cdot
\hat{\mathbf{x}}= \cos \theta_1$, $p_y\equiv \mbfp \cdot
\hat{\mathbf{y}}= \cos \theta_2$ and $p_z\equiv \mbfp \cdot
\hat{\mathbf{z}}= \cos \theta_3$, respectively. In
Fig.~\ref{fig:pxpypz_distri} we show the orientation of swimmers in
the  $p_x p_y p_z$ space for large and small reorientation times, a
particular orientation $\mbfp$ appears as a point (blue dot) on the
unit sphere. For large $\psi$, the orientation vector $\mbfp$ is
likely to explore all the possible directions, but with a slightly
marked tendency to point along $\hat{\mathbf{z}}$. In contrast, for
the small value of $\psi$, swimmers are predominantly directed along
$\hat{\mathbf{z}}$ and their spread over the unit sphere is
significantly reduced, see Fig.~\ref{fig:pxpypz_distri}. This can be understood by comparing the gyrotactic reorientation time with the characteristic timescale of the flow. For smaller values of the reorientation time ($\psi$), gyrotactic torques act more rapidly than the typical rotational fluctuations induced by the flow, leading to a more efficient alignment of the swimmers with the vertical direction ($\hat{\mathbf{z}}$). In contrast, for larger $\psi$, reorientation is slower and the flow-induced rotations allow the orientation vector to explore a broader range of directions, resulting in a nearly isotropic distribution with only a weak preferential alignment. 

We quantify this alignment tendency by computing the probability
distribution function (PDF) of $\cos \theta_3$, given that the
direction of gyrotaxis is $\hat{\mathbf{z}}$. In Fig.~\ref{fig:pdf_pz}
(a), (b) and (c), we show the PDFs $P(\cos \theta_3)$ for swimming
numbers $\phi=0$, $1$ and $10$, respectively, for different stability
numbers $\psi$ and aspect ratios $\gamma$. The three sets of color
orange, green and blue corresponds to $\psi=0.5$, $1$ and $10$,
respectively. Moreover, the three shades of the same color, in
increasing order of their lightness, correspond to three different
values of $\gamma$, i.e., spheres ($\gamma=0$, circle markers), spheroids
($\gamma=0.5$, diamond markers) and rods ($\gamma=1$, square markers). In the absence of activity,
$\phi=0$, the swimmers behave like tracers, see Fig.~\ref{fig:pdf_pz}
(a). For low stability numbers $\psi=0.5$ (orange shade curves) and
$\psi=1$ (green shade curves) swimmers show a strong alignment with
the $\hat{\mathbf{z}}$-direction, with peaks that are approximately
five times stronger than that of the latter. However, the alignment
probability decreases very rapidly as $\theta_3$ increases away from
zero. Furthermore, for a very large stability number $\psi=10$, the
distribution becomes nearly flat, suggesting that the swimmers are
able to explore all the possible directions, but the gyrotaxis along
$\hat{\mathbf{z}}$ leads to a small non-zero slope.
Figure~\ref{fig:pdf_pz} (b) shows that the qualitative behavior of the
PDFs does not change when we increase the swimming number to $\phi=1$.
Also, for these swimming numbers, we do not observe any significant
dependence of orientation on the aspect ratio.
 
Figure \ref{fig:pdf_pz}(c) shows that the tendency to align with
$\hat{\mathbf{z}}$ is enhanced for the fast swimmers ($\phi = 10$). In
agreement with this, the tails of the PDFs fall more rapidly,
suggesting a reduction in chances of anti-alignment with
$\hat{\mathbf{z}}$. Also, we observe a clear dependence on the shape
of the swimmers. We note that for small $\psi$, the
peaks of the distributions indicate that spherical and spheroidal
swimmer exhibit a marginally greater alignment tendency compared to rod-like
swimmers, whereas an opposite trend is observed for $\psi = 10$. For
the case with the largest reorientation time ($\psi=10$), the
orientation distribution is relatively flat, but the above shape
dependence is clearly evident. We observe that, as $\psi$ increases, the orientation distribution becomes broader, indicating a higher likelihood of anti-alignment among particles. For the same $\psi$, spherical swimmers show a stronger tendency toward anti-alignment compared to others. Since the swimmers mostly prefer to
orient along the vertical direction, the PDFs of the orientation along
$x$- and $y$-directions are Gaussian with zero mean (data not shown).

\begin{figure*}
	\includegraphics[width = \linewidth]{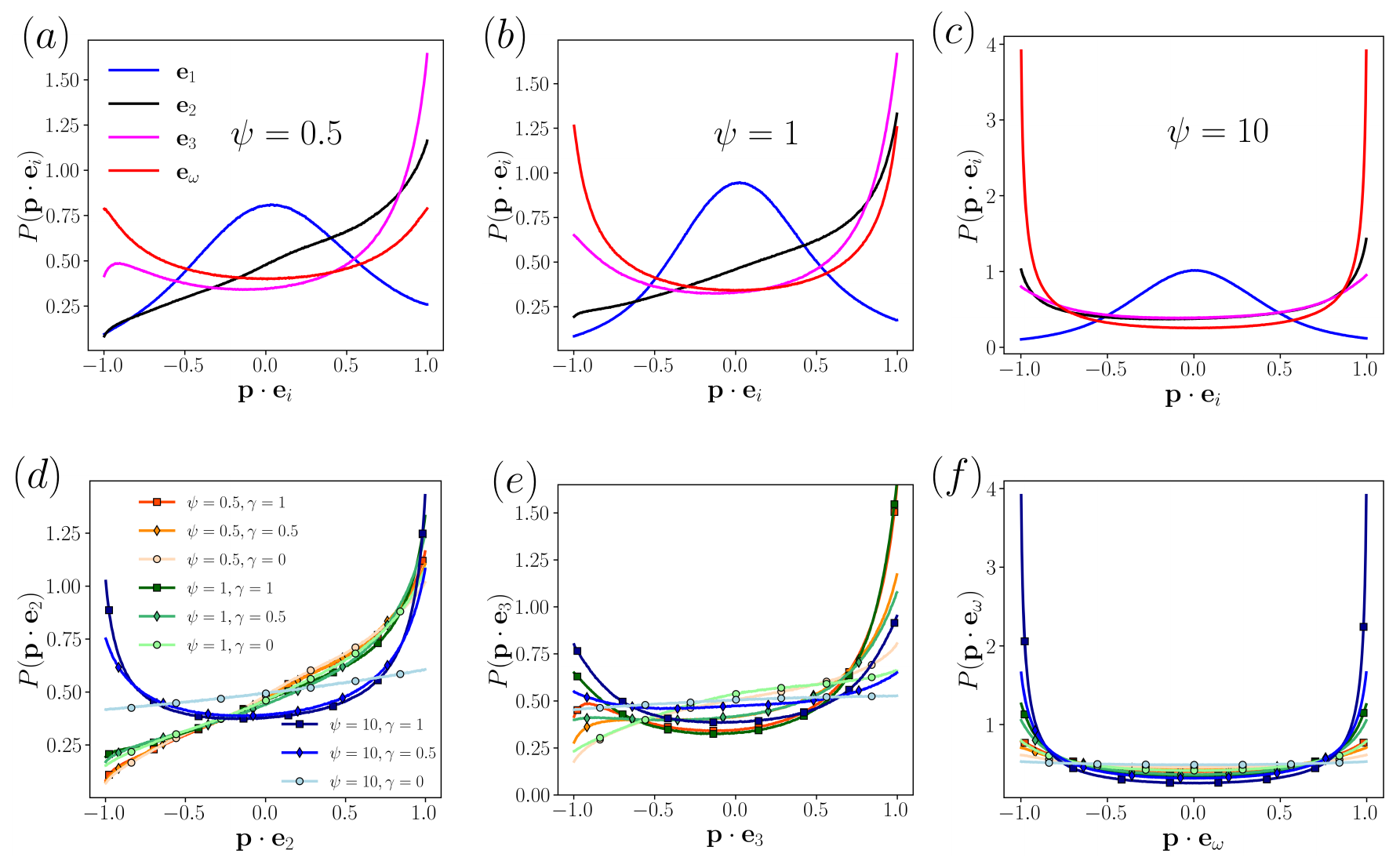}
	\vspace{-0.8cm}
	\caption{Probability distribution functions (PDFs) of the
	projections of $\mathbf{p}$  along the normalized eigenvectors of strain-rate-tensor and  the normalized vorticity for $\phi=1$ and $\gamma=1$ with reorientation time: (a) $\psi = 0.5$, (b) $\psi=1$ and (c) $\psi=10$. PDFs of $\mathbf{p}\cdot \mathbf{e}_i$ for $\phi=1$, and different values of $\psi$ and $\gamma$,  $\mathbf{e}_i$ correspond to: (d) $\mathbf{e}_2$, (e) $\mathbf{e}_3$, and (f) $\mathbf{e}_{\omega}$. The eigenvalues of the strain-rate tensor follow the order $\lambda_3 > \lambda_2 > \lambda_1$. The plots with orange, green and blue color corresponds to $\psi = 0.5$, $\psi = 1$ and $\psi = 10$, respectively. The square, diamond and circle markers corresponds to the three shapes of swimmers: rods ($\gamma=1$), spheroids ($\gamma=0.5$) and spheres ($\gamma=0$), respectively. For clarity, the legend is shown only in the first plot of each row; the same color and marker conventions apply to all other plots in the corresponding row.} 
	\label{fig:combined_pdotei}
\end{figure*}

We remark that even though the swimming number does not appear in the
governing equation for the orientation $\mbfp$, a considerable change
in the alignment of the swimmers in the vertical direction is observed
as we change $\phi$. Note that the orientation dynamics is controlled
by the effects of vorticity and strain rate in the flow, which in turn
lead to dissimilar effects on swimmers depending on their
self-propulsion speed, hence the non-trivial effect. Also, the
orientation dynamics depends on the aspect ratio, the rods experience
more shear from the flow than the spheres, see Fig.~\ref{fig:pdf_pz}
(c).

We further characterize the orientation dynamics by computing the PDFs
of the cosine of the angles that $\mbfp$ makes with the three
normalised eigenvectors of the strain rate tensor
$(\mathbf{e}_1,\mathbf{e}_2,\mathbf{e}_3)$ and the normalised local
vorticity vector $\mathbf{e}_{\omega}$ for $\phi=1$; the corresponding
eigenvalues of $(\mathbf{e}_1,\mathbf{e}_2,\mathbf{e}_3)$ satisfy
$\lambda_1 < \lambda_2 < \lambda_3$. We recall that for $\phi=1$, the
self-propulsion speed is equal to the Kolmogorov velocity. In Fig.~\ref{fig:combined_pdotei} (a),
(b) and (c), we show the PDFs of $\mbfp\cdot\mathbf{e}_i$ (where
$i=1,2,3,\omega$) for the rod-like swimmers and the stability numbers
$\psi=0.5$, $1$ and $10$, respectively. We find that the swimmers are
mostly oriented perpendicular to $\mathbf{e}_1$ (blue solid curves),
but are parallel to $\mathbf{e}_2$ and $\mathbf{e}_3$ to a varying
extent for small $\psi$ (quick reorientation time). However, the
degree of alignment or anti-alignment with $\mathbf{e}_{\omega}$
increases considerably for a large stability number $\psi=10$ (large
reorientation time). Also, we find that the PDFs of alignment with the
strain rate eigenvectors $\mathbf{e}_2$ and $\mathbf{e}_3$ are very
asymmetric for $\psi=0.5$ and $1$, this asymmetry decreases as $\psi$
increases, and the PDFs are almost symmetric for $\psi=10$.

Next, we compare the alignment of the swimmers with eigenvector
directions of the strain rate tensor and vorticity for different
values of $\gamma$ and $\psi$.  In Fig.~\ref{fig:combined_pdotei} (d),
(e) and (f) we show the PDFs of alignment of the orientation vector
with $\mathbf{e}_2$, $\mathbf{e}_3$ and $\mathbf{e}_{\omega}$,
respectively,  at $\phi=1$. Figure~\ref{fig:combined_pdotei} (d)
suggests that swimmers, irrespective of their shape and the
reorientation time, prefer to align parallel to $\mathbf{e}_2$;
however, for large reorientation time, the particle shape asymmetry
not only leads to a comparatively stronger alignment, but
significantly enhances the chances of anti-alignment. Interestingly,
Fig.~\ref{fig:combined_pdotei} (e) shows that the alignment of
swimmers with $\mathbf{e}_3$ depends considerably on the shape and the
reorientation time. Asymmetry and quick-to-moderate reorientation
times lead to stronger parallel alignment with $\mathbf{e}_3$, whereas
any decrease in asymmetry and increase in the reorientation time can
significantly lower the alignment and result in broader PDFs. In fact,
at large $\psi$ the generally favored alignment of rod-like swimmers
is comparatively reduced and leads to broader PDFs, which suggest a
reduction in preferential alignment. Finally,
Fig.~\ref{fig:combined_pdotei} (f) shows that larger reorientation
time favors an enhanced alignment or anti-alignment with
$\mathbf{e}_{\omega}$. The effect is stronger for asymmetric swimmers.
For spherically symmetric swimmers the PDFs are relatively flat,
suggesting an absence of preferential alignment.

In 3D turbulence, vortical structures are predominantly line-like or sheet-like. As a consequence of this geometry, the vorticity vector $\mathbf{e}_{\omega}$ tends to be perpendicular to the most extensional ($\mathbf{e}_3$) and most compressional ($\mathbf{e}_1$) eigenvectors of the strain-rate tensor, and exhibits a preferential alignment with the intermediate eigenvector $\mathbf{e}_2$ \cite{hamlington2008direct}. It therefore follows that the swimmers which align with $\mathbf{e}_2$ would also align with $\mathbf{e}_{\omega}$. However, we observe that swimmers with high gyrotaxis align with $\mathbf{e}_2$ and show isotropic alignment with $\mathbf{e}_{\omega}$ as seen in Fig.~\ref{fig:combined_pdotei}(d) and Fig.~\ref{fig:combined_pdotei}(f). Also, a shape dependence in the alignment statistics for low gyrotactic swimmers is observed. In particular, elongated swimmers (rods and spheroids) exhibit a tendency to both align and anti-align with $\mathbf{e}_2$ and $\mathbf{e}_{\omega}$. This is due to the fact that the swimmers reorientation time is long compared to the characteristic timescale of flow-induced rotation. As a result, the elongated swimmers are most affected by the flow gradients.

\begin{figure*}[htbp]
	\includegraphics[width = 0.6\linewidth]{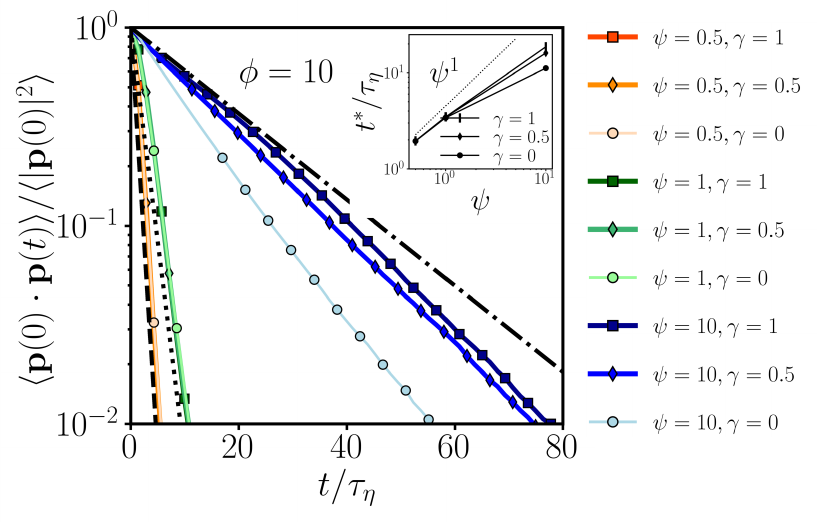}
%	\put(-220,170){\Large$\phi = 10$}
	\caption{Plots of orientation autocorrelation function $\mathcal{A}_{\mbfp}(t)$ for different aspect ratios and stability numbers at swimming number $\phi = 10$. The analytically predicted decay rate $\lambda = 1/2\psi$ is indicated by: dashed line $(\psi = 0.5)$, dotted line ($\psi=1$) and dot-dashed line ($\psi=10$). Inset: Decorrelation time vs stability number $\psi$ for rods ($\gamma=1$), spheroids ($\gamma=0.5$) and spheres ($\gamma=0$). The plots with orange, green and blue color corresponds to $\psi = 0.5$, $\psi = 1$ and $\psi = 10$, respectively. The square, diamond and circle markers corresponds to the three shapes of swimmers: rods ($\gamma=1$), spheroids ($\gamma=0.5$) and spheres ($\gamma=0$), respectively.} 
	\label{fig:auto_corr_phi}
\end{figure*}

\subsection{Orientation autocorrelation}
We define the orientation autocorrelation as 
\begin{equation}
\mathcal{A_{\bf{p}}} = \frac{\langle {\bf{p}}(0) \cdot {\bf{p}}(t) \rangle}{\langle {|\bf{p}}(0)|^2 \rangle},
\label{eqn:auto_corr}
\end{equation}
where $\mbfp(0)$ is the orientation vector at $t=0$. In Fig.~\ref{fig:auto_corr_phi} we show $\mathcal{A_{\bf{p}}}$ vs $t/\tau_{\eta}$ for different shapes and reorientation times at a fixed swimming number $\phi=10$. The autocorrelation decays exponentially with a decay rate that depends strongly on the stability number $\psi$. For $\psi=10$ the decay occurs over a long period $t \approx 75 \tau_{\eta}$, whereas for $\psi=0.5$ and $1$ the decay period is significantly short $t \leq 10$. We find that the orientation autocorrelation for fast swimmers decays as $\mathcal{A}_{\bf{p}} \propto C_0 e^{-\lambda\frac{ t}{\tau_{\eta}}}$, with $\lambda = 1/2\psi$. We estimate the decorrelation time $t^*$ as the interval over which $\mathcal{A}_{\bf{p}}(t^*) = \mathcal{A}_{\bf{p}}(0)/e$ and show its variation with $\psi$ in the inset of Fig.~\ref{fig:auto_corr_phi} for three different shapes. We observe a deviation from linearity at large $\psi$, which is more pronounced for the spheres. Note that spherical swimmers with large reorientation time constitute an exception, as $\mathcal{A}_{\bf{p}}$ decays with rate $\lambda \approx 1/\psi$. For $\phi=0$ and $1$, the autocorrelation drops to zero over relatively shorter intervals $6 \tau_{\eta}$, $10 \tau_{\eta}$ and $25 \tau_{\eta}$ for $\psi=0.5$, $1$ and $10$, respectively (data not shown). Clearly, swimmers with quick reorientation time de-correlate fast from their random initial alignment, indicating a quick memory loss of their past states.

We can understand the above exponential behavior by considering the time derivative of Eq.~\eqref{eqn:auto_corr} and using Eq.~\eqref{eqn:pdot} to obtain
\begin{align}
	\frac{d \mathcal{A_{\bf{p}}}}{dt} = \frac{1}{\langle|\mbfp(0)|^2\rangle} \Bigg\langle 
	\mbfp(0) \cdot  \frac{\left[\hat{\mathbf{z}} - (\hat{\mathbf{z}}\cdot \mbfp)\mbfp \right]}{2B} + \mbfp(0) \cdot \frac{\boldsymbol{\omega} \times \mbfp}{2} \nonumber 
	+ \mbfp(0) \cdot \gamma \left[\mathbf{S}\mbfp - (\mbfp \cdot \mathbf{S}\mbfp)\mbfp \right] 
	\Bigg\rangle.
	\label{eqn:orient_1}
\end{align}
Figure~\ref{fig:auto_corr_phi} suggests that $\mathcal{A_{\bf{p}}}$ is
almost independent of the aspect ratio for any given $\psi$ in our DNS
results. Hence, we neglect the last term in the above equation. Also,
as the reorientation time is smaller than the vorticity time scale for
low values of $\psi$, the gyrotaxis term dominates. Therefore, the
second term can be neglected. The term $\langle \mbfp(0) \cdot
\hat{\mathbf{z}} \rangle $ averages to zero, and we make a reasonable
assumption that the swimmers with low $\psi$ are well aligned with the
vertical direction, $(\hat{\mathbf{z}} \cdot \mbfp)$ approaches unity.
These along with the definition $\psi=B/\tau_{\eta}$ give $d
\mathcal{A_{\bf{p}}}/dt \approx -(1/2\psi\tau_{\eta})
\mathcal{A_{\bf{p}}}$, whose solution is $\mathcal{A_{\bf{p}}} \approx
C_0 e^{-\frac{1}{2 \psi} \frac{t}{\tau_{\eta}}}$. This suggests that
the gyrotactic terms play the dominant role in the decorrelation of
swimmer orientation.
 
%%%%%%%%%%%%%%%%%%%%%%%%%%%%%%%%%%%%%%%%%%%%%%%%%%%%%%%%%%%%%%%%%%
\subsection{Transport}
\begin{figure*}
	\includegraphics[width = \linewidth]{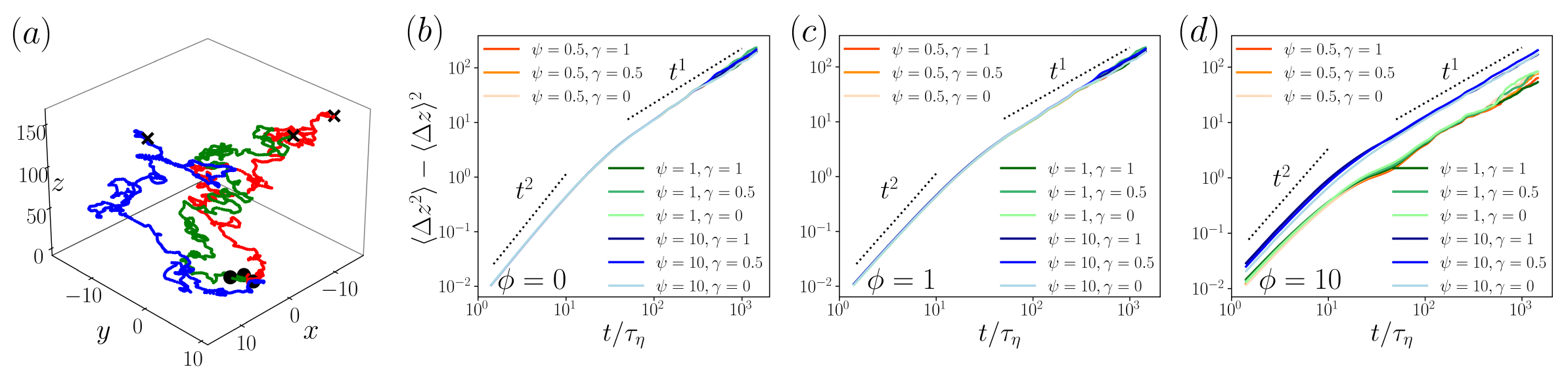}
	\caption{(a) Trajectories of three randomly chosen swimmers, where
	\protect\disc\ and \protect\xmarker\ mark the starting and ending
	positions, respectively. Mean-squared displacement along the vertical direction
	for swimming numbers: (b) $\phi = 0$, (c) $\phi = 1$, and (d) $\phi
	= 10$. For each $\phi$, following combinations of $\psi=(0.5,1,10)$
	and aspect ratios $\gamma=(1,0.5,0)$ were considered.} 
	\label{fig:zmsd_phi}
\end{figure*}
We examine the transport of microswimmers by computing the mean-squared displacement (MSD)
\begin{equation}
\langle \delta \mathbf{r}^2 \rangle = \left\langle \frac{1}{N} \sum_{i=1}^{N} |\mathbf{r}_i(t) - \mathbf{r}_i(0)|^2 \right\rangle,
\label{eqn:msd}
\end{equation}
where $\mbfr_i(0)$ is the initial position. The gyrotactic microswimmers tend to align along the vertical direction, so they spend a considerable time traveling in the vertical direction compared to the horizontal direction. Figure~\ref{fig:zmsd_phi} (a) shows the expected comparative extent of the trajectories along the $x$-, $y$- and $z$-directions, microswimmers are predominantly displaced along the $z$-direction. Therefore, while accounting for this bias, we compute the MSD of the swimmers in the z-direction as
\begin{equation}
\langle \Delta z^2 \rangle = \langle \delta z(t)^2 \rangle - \langle \delta z(t) \rangle^2 , 
\label{eqn:msd}
\end{equation}
 where $\delta z = |z_i(t) - z_i(0)|$. Figure~\ref{fig:zmsd_phi} (b), (c) and (d) show the MSD $\langle \Delta z^2 \rangle$ for swimming numbers $\phi=0$, $1$ and $10$, respectively. These plots unequivocally show that the swimmers, irrespective of their shape and stability numbers, exhibit a ballistic motion $\langle \Delta z^2 \rangle \propto t^2$ at short intervals and a diffusive behavior at long intervals $\langle \Delta z^2 \rangle \propto t$.

%\subsection{Trajectory Analysis}
\begin{figure*}
	\includegraphics[width = \linewidth]{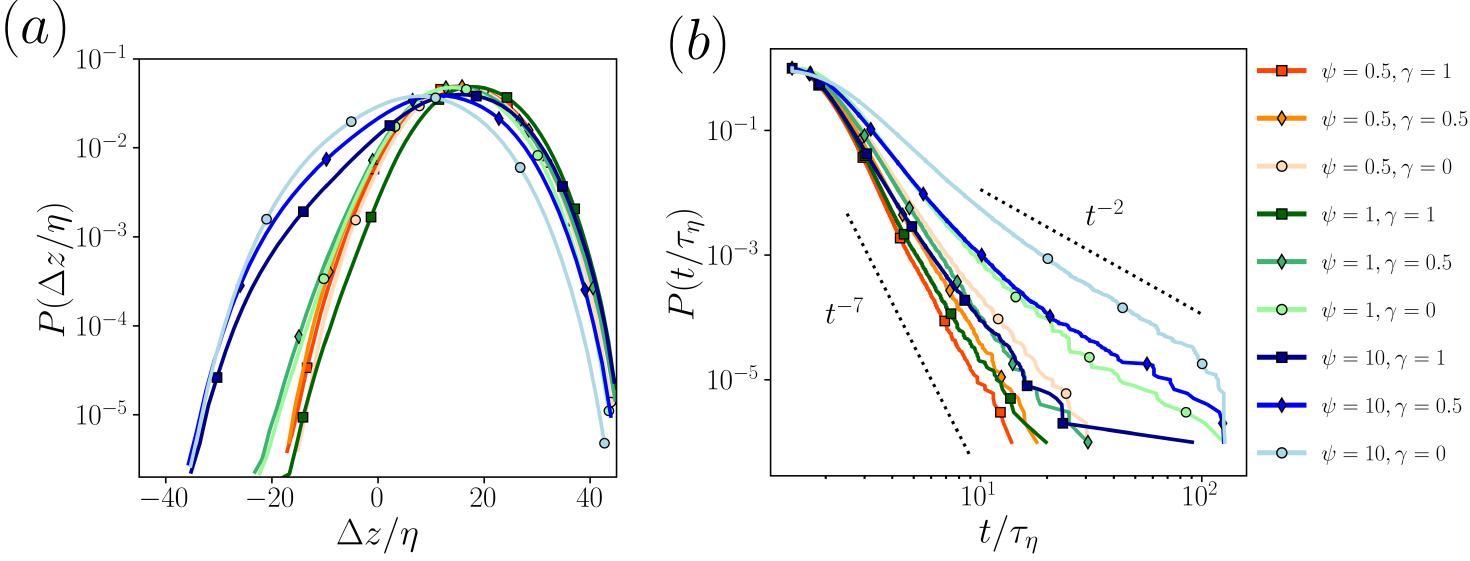}
%	\put(-510, 180){\large (a)}
%	\put(-265, 180){\large (b)}
		\caption{Probability distribution functions of: (a) the vertical
		displacement $\Delta z$ in a fixed time interval of $1.4
		\tau_{\eta}$, (b) the time taken to cover a fixed vertical
		displacement of $1000\eta$. The swimming number $\phi=10$ is held
		fixed, while the combinations of $\psi=(0.5,1,10)$ and
		$\gamma=(1,0.5,0)$ were considered. The plots with orange, green and blue color corresponds to $\psi = 0.5$, $\psi = 1$ and $\psi = 10$, respectively. The square, diamond and circle markers corresponds to the three shapes of swimmers: rods ($\gamma=1$), spheroids ($\gamma=0.5$) and spheres ($\gamma=0$), respectively.} 
	\label{fig:z_distance_time}
\end{figure*}

To further characterize the transport of microswimmers, we compute the distribution of vertical displacement in a given time interval. In Fig.~\ref{fig:z_distance_time} (a), we show PDFs for the fixed time interval $1.4 \tau_{\eta}$ and $\phi=10$. We find that the distributions, in general, are asymmetric. For small stability numbers (fast reorientation times), for example, $\psi=0.5$ and $1$, the observed distributions are narrow with a larger net positive displacement, the PDFs peak at $\Delta z = 20\eta$. However, the left tail of the PDF significantly broadens as $\psi$ is increased to $10$, now covering nearly equal ranges of positive and negative displacements. Also, the dependence on the aspect ratio is either relatively weak or altogether absent. Moreover, for reduced swimming numbers, both the extent of the distribution as well as the mean displacement are reduced. In Fig.~\ref{fig:z_distance_time} (b), we show the distribution of the time taken to cover a fixed vertical distance $1000\eta$ for $\phi=10$. The tails of the PDFs $P(t/\tau_{\eta})$ exhibit power laws $t^{-\beta}$, with exponents that strongly depend on the aspect ratio. As $\psi$ decreases, the exponent $\beta$ increases from $2$ to $7$. Also, for a given $\psi$, the exponents order as $\beta_{\text{rods}} > \beta_{\text{spheroids}} > \beta_{\text{sphere}}$. This is consistent with the findings of \cite{lovecchio2019chain}, which show that chain formation can enhance the vertical migration rate of phytoplankton.

\section{Reduced Model}
We use a reduced model to capture essential features of the orientation dynamics obtained using the DNS run. It  qualitatively interprets the DNS results using a minimal stochastic model, where the turbulent velocity field is represented by Gaussian random noise, capturing the essential trends without the complexity of fully resolved turbulence. For simplicity, we consider a 2D model and focus only on spherical microswimmers $\gamma=0$, wherein gyrotaxis occurs along the $z$-axis \cite{fouxon2015phytoplankton, santamaria2014gyrotactic}. We replace the background flow velocity $\mbfu(\mbfx,t)$ in Eq.~\eqref{eqn:xdot} and the vorticity in Eq.~\eqref{eqn:pdot} by Gaussian random noise $\xi \equiv (\xi_x, \xi_z)$ and $\xi_{\theta}$, respectively. We prescribe $\langle \xi_i(t) \rangle = 0$ and $\langle \xi_i(t) \xi_j(t') \rangle = \sigma_i^2\delta_{ij} \delta(t-t')$, where $\sigma_i$ is the strength of the noise. Equations describing the evolution of the position and the orientation vector $\mbfp\equiv (\cos \theta, \sin \theta)$ are given by
\begin{subequations}
 \begin{gather} 		
 		\dot{x} = v_s \cos{\theta} +  \xi_x,  \label{eqn:xdot_model}\\
	    \dot{z} = v_s \sin{\theta} +  \xi_z, \label{eqn:ydot_model}\\
       \dot{\theta} = \frac{\cos{\theta}}{2 B} + \frac{1}{2} \xi_{\theta}.
       \label{eqn:thetadot_model}
 \end{gather}
\end{subequations}
We define the swimming number as $\phi=v_s/u_{x\theta}$ and the stability number as $\psi = B \sigma_{\theta}^2$, where $u_{x\theta}=\sigma_x\sigma_{\theta}$ and $u_{z\theta}=\sigma_z \sigma_{\theta}$. 

\begin{figure*}
	\includegraphics[width = 0.95\linewidth]{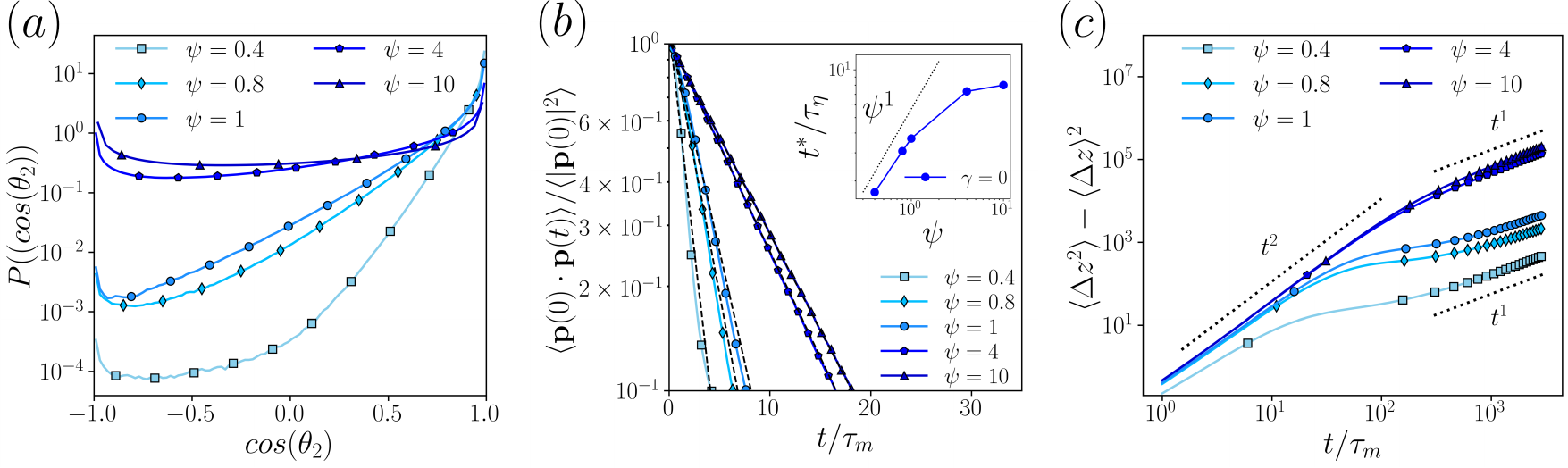}
	\label{fig:msd_model}
%	\put(-475, 130){\large (a)}
%	\put(-320, 130){\large (b)}
%	\put(-158, 130){\large (c)}
	\caption{Reduced model for spherical microswimmers. (a) Probability distribution of orientation about the $z$-axis $P(\cos \theta_{2})$, (b) Orientation autocorrelation and (c) mean-squared displacement  along the $z$-axis for the stability numbers $\psi=(0.4,0.8,2,4,10)$, while the swimming number is held fixed at $\phi=10$.} 
	\label{fig:pdf_msd_reduced_model}
\end{figure*}

We perform numerical simulations of the above equations for several values of swimming and stability numbers to explore the statistics of orientation dynamics. In Fig.~\ref{fig:pdf_msd_reduced_model} (a) we show the PDFs of orientation of swimmers along the $z$-direction, $i.e.,$ $P(\cos \theta_2)$ for different values of $\psi$ at fixed $\phi=10$. We observe that the general features are similar to those observed in the DNS run. For quick reorientation time almost all the particles are oriented along the $z$-direction. However, as we increase the reorientation time (larger $\psi$), the swimmers begin to explore directions away from the $z$-axis; the tails of the PDFs become broader and fatter. A similar trend is observed for other values of the swimming number (data not shown). In Fig.~\ref{fig:pdf_msd_reduced_model} (b), we show the orientation autocorrelation functions for different values of $\psi$. Once again, in agreement with the DNS results, we find that $\mathcal{A_{\bf{p}}}$ decays exponentially with time and the decay rates are given by $\lambda_{model} \approx 1/2\psi$. Inset shows the linear dependence of decorrelation time with reorientation time with a small deviation at larger $\psi$, similar to the DNS results. Figure~\ref{fig:pdf_msd_reduced_model} (c) shows that the MSD displays a ballistic transport at short time intervals and a diffusive behavior at large time intervals, which is in agreement with DNS results. However, we notice that the time scale associated with the crossover from the ballistic to diffusive regime increases with $\psi$. 

\begin{figure*}
	\includegraphics[width = 0.6\linewidth]{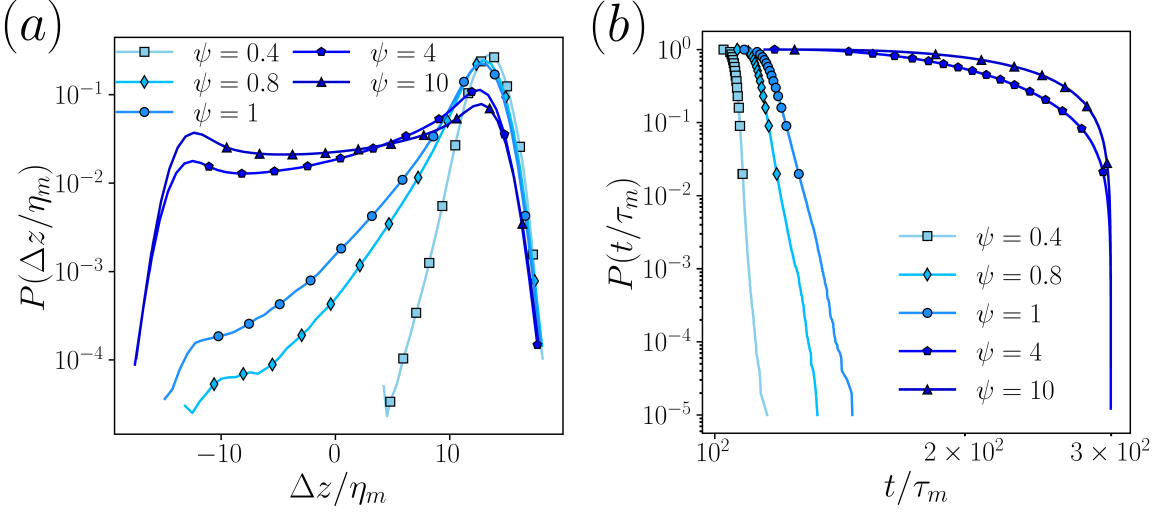}
	\label{fig:msd_model}
%	\put(-298, 130){\large (a)}
%	\put(-145, 130){\large (b)}

	\caption{Reduced model for spherical microswimmers. Probability distribution functions of: (a) the vertical displacement $\Delta z$ in a fixed time interval of $1.4 \tau_{m}$, (b) the time taken to cover a fixed vertical displacement of $1000\eta_m$ for the stability numbers $\psi=(0.4,0.8,2,4,10)$, while the swimming number is held fixed at $\phi=10$. $\tau_{m}=1/\sigma^2_{\theta}$ and $\eta_m=\sigma_x/\sigma_{\theta}$, where $\sigma_x$ and $\sigma_{\theta}$ are the standard deviations of translational and rotational noise, respectively.} 
	\label{fig:pdf_y_distance_time}
\end{figure*}

In Fig.~\ref{fig:pdf_y_distance_time} (a) we show the PDFs of displacement along the $z$-direction in a fixed interval of time $1.4 \tau_{m}$, where $\tau_{m}=1/\sigma^2_{\theta}$. We find that the PDFs are qualitatively similar to those obtained from the DNS run. For small values of $\psi$, the PDFs are narrow and only positive displacements are observed. However, when $\psi$ is increased, the PDFs become asymmetric and the left tail becomes broad. Figure~\ref{fig:pdf_y_distance_time} (b) shows the PDFs of time taken to cover a fixed vertical displacement $1000\eta_m$, where $\eta_m=\sigma_x/\sigma_{\theta}$. The behavior is qualitatively similar to what we observe in DNSs, as we increase $\psi$ the PDFs fall less rapidly and become fatter.

\section{Conclusions}

We examined the orientation statistics of spherical, spheroidal and
rod-like gyrotactic microswimmers in a turbulent flow. The probability
distribution of the vertical alignment of swimmers exhibits a clear
dependence on the gyrotaxis strength. The swimmers with higher
gyrotactic strength tend to align more along the vertical direction
compared to those with lower gyrotactic strength. Also, for low
swimming numbers ($\phi \le 1$), the orientation distribution is nearly
independent of the aspect ratio ($\gamma$) considered. However, for
large swimming numbers ($\phi=10$), spheres and spheroids show marginally greater
alignment with the vertical compared to rods at low $\psi$, but the
distribution is nearly flat for high $\psi$ with rod-like swimmers
showing marginally higher alignment with the vertical.
 
To further understand this behavior, we examined the alignment with
the strain-rate-tensor eigenvectors and the vorticity vector. Our
results show that rod-like swimmers with high gyrotactic strength
preferentially align with the eigenvector corresponding to the largest
strain-rate eigenvalue (also known as the Lagrangian stretching
direction), while those with low gyrotactic strength align more with
the vorticity vector. This contrasts with the behavior of
non-gyrotactic swimmers, where rods tend to align more with vorticity
than with strain-rate eigenvectors~\cite{zhan2014accumulation,
pumir2011orientation}. Here, we systematically compared the influence
of three aspect ratios under
identical turbulent and gyrotactic conditions, revealing how geometry
and gyrotaxis jointly govern the alignment. We note that our results
are for a modest value of $Re_{\lambda} \approx 120$. However, the
Reynolds number, along with $\gamma \in [-1,1]$, can influence the
angle between the swimming direction and the flow velocity
$\cos(\theta_u) \propto \gamma
Re_{\lambda}^{-1/2}$~~\cite{borgnino2019alignment}. Similarly,
preferential sampling in the vertical direction depends on both
$Re_{\lambda}$ and $\psi$~\cite{borgnino2018gyrotactic}. While higher
Reynolds numbers at fixed $\psi$ enhance sampling in downwelling
regions, increasing $\psi$ reduces this tendency, causing swimmers to
sample upwelling regions more frequently.

Our numerical results, along with analytical estimates, show that the
orientation autocorrelation decays exponentially with rate $1/2\psi$
at high activity, mostly irrespective of the shape. However, we observe a deviation from this for spherical
swimmers at large $\psi$. For axisymmetric non-gyrotactic tracers, the
decorrelation time is known to exhibit a slight increase with Reynolds
number~\cite{pumir2011orientation}. Also, we find that irrespective of
their shape, gyrotactic response and swimming number, swimmers exhibit
a ballistic and a diffusive transport behavior in the short- and
long-time limit, respectively. Moreover, the PDFs of time taken to
cover a specified displacement exhibit power-law tails, whose
exponents may depend on the gyrotactic response and the shape of
microswimmers. A two-dimensional minimal model, wherein the flow
velocity and its gradients are replaced by the Gaussian white noise, is able to capture
the essential features of the orientation dynamics observed in the DNS
run.

The present results have important implications for the ecology of motile microorganisms in natural aquatic environments. The alignment of elongated swimmers influences light penetration and scattering in the ocean, thereby affecting photosynthesis \cite{marcos2011microbial}. In addition, vertical migration plays a key role in regulating access to light and nutrients for phytoplankton and other microswimmers \cite{durham2013turbulence, hamlington2008direct}. Our results indicate that elongated swimmers exhibit enhanced vertical migration, consistent with \cite{lovecchio2019chain}, which shows that chain formation can increase migration rates. In regions of high shear, rod-like swimmers are more susceptible to reorientation, leading to increased patchiness and the formation of thin phytoplankton layers. Such structuring can strongly influence collective behavior, promoting processes such as reproduction and competition, while also impacting light penetration; in the case of toxic blooms, it may even contribute to zooplankton and fish mortality \cite{santamaria2014gyrotactic}.

\begin{acknowledgements}
The authors acknowledge National Supercomputing Mission
(NSM) for providing computing resources of ``PARAM Shakti'' at IIT Kharagpur and ‘PARAM Porul’ at NIT Trichy, which is implemented by C-DAC and supported by the Ministry of Electronics and Information Technology (MeitY)
and Department of Science and Technology (DST), along with
the NSM Grant DST/NSM/R$\&$D HPC Applications/2021/03.21. VS
would like to acknowledge support from the Institute Scheme for
Innovative Research and Development (ISIRD), IIT Kharagpur,
Grant No. IIT/SRIC/ISIRD/2021-2022/03. SKN thanks the Prime Minister's Research Fellows (PMRF) scheme, Ministry of Education, Government of India. 

\end{acknowledgements}
%\bibliographystyle{apsrev4-2}
%\bibliography{reference}

\end{document}